\documentclass[prodmode,acmtap]{ci2018}

\acmVolume{2}
\acmNumber{3}
\acmArticle{1}
\articleSeq{1}
\acmYear{2010}
\acmMonth{5}

\usepackage[ruled]{algorithm2e}
\SetAlFnt{\algofont}
\SetAlCapFnt{\algofont}
\SetAlCapNameFnt{\algofont}
\SetAlCapHSkip{0pt}
\IncMargin{-\parindent}
\renewcommand{\algorithmcfname}{ALGORITHM}

\title{CrowdRev: A platform for Crowd-based Screening of Literature Reviews}
\author{JORGE RAMIREZ and EVGENY KRIVOSHEEV \affil{University of Trento}
MARCOS BAEZ and FABIO CASATI
\affil{University of Trento and Tomsk Polytechnic University}
BOUALEM BENATALLAH
\affil{University of New South Wales}
}

\begin{abstract}
abstract
\end{abstract}

\begin{document}

\maketitle

\section{Background and Motivation}




This paper presents a process and system - along with a demo flow - for crowd-supported \textit{systematic literature reviews} (SLRs). 
SLRs in general and meta analyses in particular are one of the most important kind of scientific publication: they summarize results achieved by different research groups and, thanks to a systematic process ~\cite{Grant2009ReviewTypes,khan2003five,henderson2010write}, aimed at finding many relevant papers and selecting them in an unbiased and transparent way, they are supposed to provide a comprehensive and objective summary of research in a field ~\cite{steinberg2011clinical,Haidich2010}. For this reason, especially in medicine, they form the basis of policy decisions. They are also one of the most highly cited form of publication, and their number is growing steadily in all fields of science \cite{Krivosheev_hcomp17}.

The commonly accepted SLR process starts with defining and executing a keyword-based boolean query over paper databases such as Scopus, typically returning thousands of results. 
For example, for an SLR that reports on studies on how technology can help social interactions among older adults, we may query for papers that include in the title or abstract keywords such as \textit{technology}, \textit{internet}, \textit{intervention}, \textit{study}, \textit{older adults}, etc.  
The results is likely to include many irrelevant papers that authors filter out based on a set of exclusion criteria, such as ``does the paper describe an intervention-type study", or ``does the paper describes a study involving 75+ older adults". 
This is typically done based on title and abstract only. 
Papers that survives the screening (often a small fraction) are then read in full and, if still considered relevant, analyzed and discussed in the SLR.

As it can be appreciated, performing SLRs is not easy: the process is long, laborious,  error prone, and possibly frustrating.
As we witnessed, sometimes the process becomes so long that the work is not completed and not made available in a form useful for the community to consume.
This also applies to the screening phase, often lasting for several months. 
The fact that literally millions of scientific papers are published every year makes the task more daunting~\cite{coch_handbook2011,Sampson2008,Krivosheev_hcomp17}.
Even when completed and published, it has been shown that it is not uncommon for SLRs to miss a very high percentage of relevant papers (well above 30\% ~\cite{wasted2016}).

In this paper and demo we present a crowd and crowd+AI based system, called CrowdRev, supporting the screening phase of literature reviews and achieving the same quality as author classification at a fraction of the cost, and near-instantly. CrowdRev makes it easy for authors to leverage the crowd, and ensures that no money is wasted even in the face of difficult papers or criteria: if the system detects that the task is too hard for the crowd, it just gives up trying (for that paper, or for that criteria, or altogether), without wasting money and never compromising on quality. 
The theory for this work is laid out in our recent publications \cite{Krivosheev_hcomp17,Krivosheev_www18} (also showing experimental results) and is inspired by a number of contributions by many authors that have shown the feasibility of crowd-based screening of scientific papers, even in complex fields such as medicine~\cite{Wallace2017crowdML,Sun16-hcomp,Mortensen2016crowd,Weiss2016}.

\begin{figure*}
\centering
\includegraphics[width=\textwidth ]{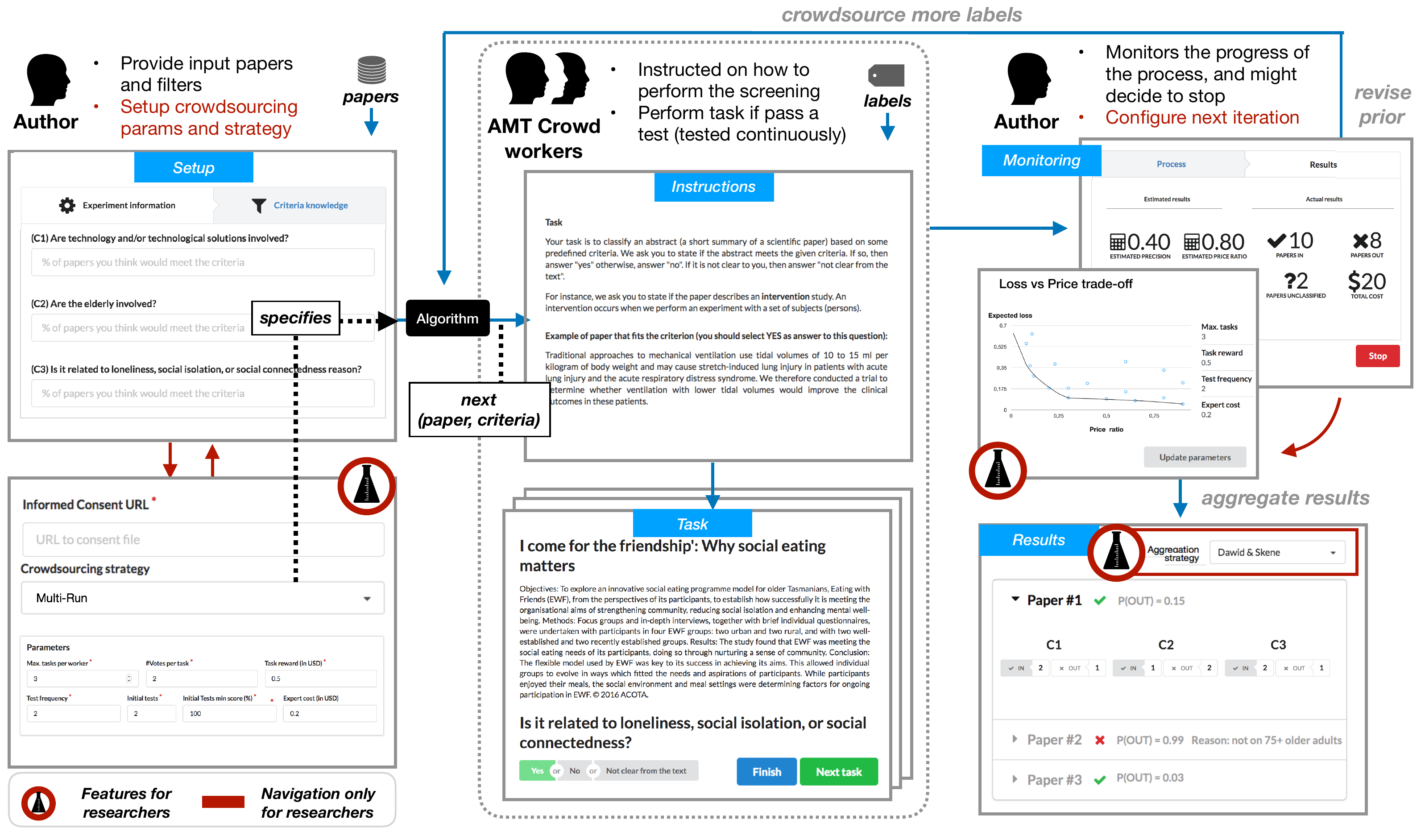}
\caption{Example screening workflow}
\label{fig:flow}
\end{figure*}


\section{Process and Interaction Flow}
CrowdRev has two interaction modes, depending on users and their needs: i) for authors who wants to screen papers and do not care to deal with any of the intricacies of crowdsourcing, and ii) for crowdsourcing researchers, who want to experiment with crowd and hybrid crowd-machine classification.

In the author flow, the interaction with the system proceeds as follows.
Authors upload the exclusion criteria, the list of candidate papers (the result of the initial query on paper databases), and a set of 10 or more test items (papers for which the true label is known).
CrowdRev prepares the training material for each criteria (a screen where a positive and negative example for each criterion are shown to the crowd workers, taken from test items). 
Authors can also see historical cost estimates, so they can get an idea of how much an average screening in their field might cost. 

The crowd classification then begins. Workers are selected from Mechanical Turk (AMT): the platform generates a Human Intelligence Tasks (HIT) on AMT that refers workers to CrowdRev. 
Each worker is assigned a criteria, and classifies papers for that criteria as discussed later.
Fig \ref{fig:flow}(middle) shows what the worker sees as task. 
We select workers that either have accuracy over 70\% or that are qualified for SLR specifically, based on badges we assign. 
Two initial tests and then periodic ``honeypots" tests are inserted to screen inaccurate workers.

The power of the tool lies in optimizing which (paper,criterion) pairs to obtain votes for. 
We have two objectives here: first, for each paper $p$, we want to identify a criterion $c$ that is likely to apply to $p$ (if at least \textit{one} criterion applies to $p$, it is screened out, we do not need to assess the other criteria) and that is not \textit{difficult} for the crowd to assess, that is, we can trust crowd votes and therefore reach a decision on classification with fewer votes. Second, we want to identify papers for which we are not likely to reach an accurate decision cheaply: for these papers, we just give up and leave them to authors to classify, without wasting additional money.
This makes the tool robust to difficult problems, meaning that CrowdRev can always be used: in the worst case, it will give up soon, spending only a minimal amount.

CrowdRev achieves this by leveraging a crowd- and crowd+AI-based screening algorithm we developed, called Shortest Run in the following (SR) \cite{Krivosheev_www18}. 
In essence, SR works by computing the probability that each criterion $c$ applies to a paper $p$, based on current votes, and on a prior probability which is the proportion of papers to which that criterion applies. 
Based on this, through simple Bayesian statistics and through estimates of the crowd accuracy in assessing $c$, we can also estimate i) the probability $P^{out}_{p,c}$ that criteria $c$ applies to paper $p$ (and in general, via simple math, the probability that $P^{out}_p$ that paper $p$ should be screened out, which we do if $P^{out}_p > 0.99$), 
ii) the probability of obtaining \textit{out} votes if we query the crowd for $(p,c)$, and iii) the probability that we will reach a decision after receiving few out votes. 
We can then decides the pairs $(p,c)$ we query the crowd for by prioritizing the pairs that will allow us to reach a decision cheaply.

Authors can launch an entire classification or can proceed in arbitrarily small steps, investing a small amount each time. 
The only exception is the initial run for which we ask at least 3 votes per paper and per criteria for 20 papers (for a total cost of about 6\$ per criterion). 
SR needs this for the initial estimation of criteria power and difficulty statistics, which in turn enables us to estimate the classification accuracy and cost. 
Estimates and results are updated as the classification progresses. The process ends either when the authors decide to stop or when SR believes it has classified all it could do so cheaply. 
The process happens in near real time: crowd classification completes within a few hours, and at a fraction of the author cost \cite{Krivosheev_www18}.  
Notice also that the process is transparent, and the detailed results of the screening process can be added as supplementary material to the SLR, with the precise definition of criteria and the results of the decision process. 
This is something that readers can very rarely see in SLR papers today.

The researcher (or advanced user) process flow is different (items labeled via the red circle in Figure \ref{fig:flow}), in that users can select i) which strategy to adopt \textit{before} polling the crowd (to reduce the number of votes we ask), and ii) which crowdsourced classification (label aggregation) method to leverage \textit{after} the votes have been obtained, along with other parameters.
In Shortest Run the two are coupled, as the aggregation method determines the progression of the algorithm in terms of which votes to ask next, but for other algorithms (such as the most common one where $J$ votes per paper and per criterion are asked from the crowd), users can experiment with a variety of vote aggregation algorithms, such as ~\cite{DawidSkene_Confusion,whitehill2009whose,Dong2013,Li_error_13,LiuWang_truelabel,liu2013scoring,karger2011iterative,Zhou_spectral_13,Ok_belief_13}.
In this mode users can also view a curve that indicates the expected recall or precision depending on the budget and algorithm chosen: in general, the higher the budget, the more votes we can ask, and the more precise the classification is.
The tool does this via simulations, first based on generic, historical estimates (leading to  priors with little informative value) and by then refining the estimates at the classification process begins - thereby also giving users the possibility to adjust their preferences. The tool always also identifies the pareto-optimal curve, in which the points may belong to different algorithms.
Additional information, data, experimental results, as well as the tool in alpha version is available at http://jointresearch.net.




\newpage

\bibliographystyle{ci-format}
\bibliography{cs}

\end{document}